\documentclass[twocolumn,secnumarabic,amssymb,nobibnote,aps,prl,superscriptaddress]{revtex4-1}
\pdfoutput=1
\usepackage{graphicx} 
\usepackage{hyperref}
\usepackage[english]{babel}
\usepackage{datetime}
\usepackage{verbatim}
\usepackage{amsmath}
\usepackage{amsfonts}
\usepackage{amssymb}
\usepackage{siunitx} 
\usepackage{nccmath}
\usepackage{dcolumn}
\usepackage{lineno}
\usepackage{xcolor}
\usepackage{color}
\usepackage{bm}
\usepackage{flexisym}
\usepackage{booktabs}
\usepackage{longtable}
\usepackage{epstopdf}
\usepackage{wrapfig}
\usepackage[utf8]{inputenc}
\usepackage{comment}
\usepackage{lipsum} 
\usepackage[normalem]{ulem}
\usepackage{lmodern}

\hypersetup{colorlinks,bookmarksopen,bookmarksnumbered,
citecolor=blue,
linkcolor=black,
pdfstartview=green,
urlcolor=blue}

\addto\captionsenglish{}

\newcommand{\Eqref}[1]{Eq.~\eqref{#1}}
\newcommand{\Fref}[1]{Fig.~\ref{#1}}

\def\surfT{\gamma}    
\def\young{Y}         
\def\bend{\kappa}
\def\dens{g\Delta\rho}
\def\fvk{\Upsilon}    
\def\dimSurf{\Gamma}  
\def\squash{\Pi}      

\makeatletter
\let\cat@comma@active\@empty
\makeatother

\begin{document}
\title{Faceting and flattening of emulsion droplets: a mechanical model}

\author{Ireth Garc{\'i}a-Aguilar}
\affiliation{Instituut-Lorentz, Universiteit Leiden, P.O. Box 9506, 2300 RA Leiden, Netherlands}
\author{Piermarco Fonda}
\affiliation{Instituut-Lorentz, Universiteit Leiden, P.O. Box 9506, 2300 RA Leiden, Netherlands}
\affiliation{Theory \& Bio-Systems, Max Planck Institute of Colloids and Interfaces, Am M\"uhlenberg 1, 14476 Potsdam, Germany}
\author{Eli Sloutskin}
\affiliation{Physics Department and Institute of Nanotechnology and Advanced Materials, Bar-Ilan University, Ramat Gan 529002, Israel}
\author{Luca Giomi}
\affiliation{Instituut-Lorentz, Universiteit Leiden, P.O. Box 9506, 2300 RA Leiden, Netherlands}
\email{giomi@lorentz.leidenuniv.nl}

\date{\today}

\begin{abstract}
When cooled down, emulsion droplets stabilized by a frozen interface of alkane molecules and surfactants have been observed to undergo a spectacular sequence of morphological transformations: from spheres to faceted liquid icosahedra, down to flattened liquid platelets. While generally ascribed to the interplay between the elasticity of the frozen interface and surface tension, the physical mechanisms underpinning these transitions have remained elusive, despite different theoretical pictures having been proposed in recent years. In this article, we introduce a comprehensive mechanical model of morphing emulsion droplets, which quantitatively accounts for various experimental observations, including the size scaling behavior of the faceting transition. Our analysis highlights the role of gravity and the spontaneous curvature of the frozen interface in determining the specific transition pathway.
\end{abstract}

\maketitle

\begin{figure*}[t!]
    \centering
    \includegraphics[width=0.95\textwidth]{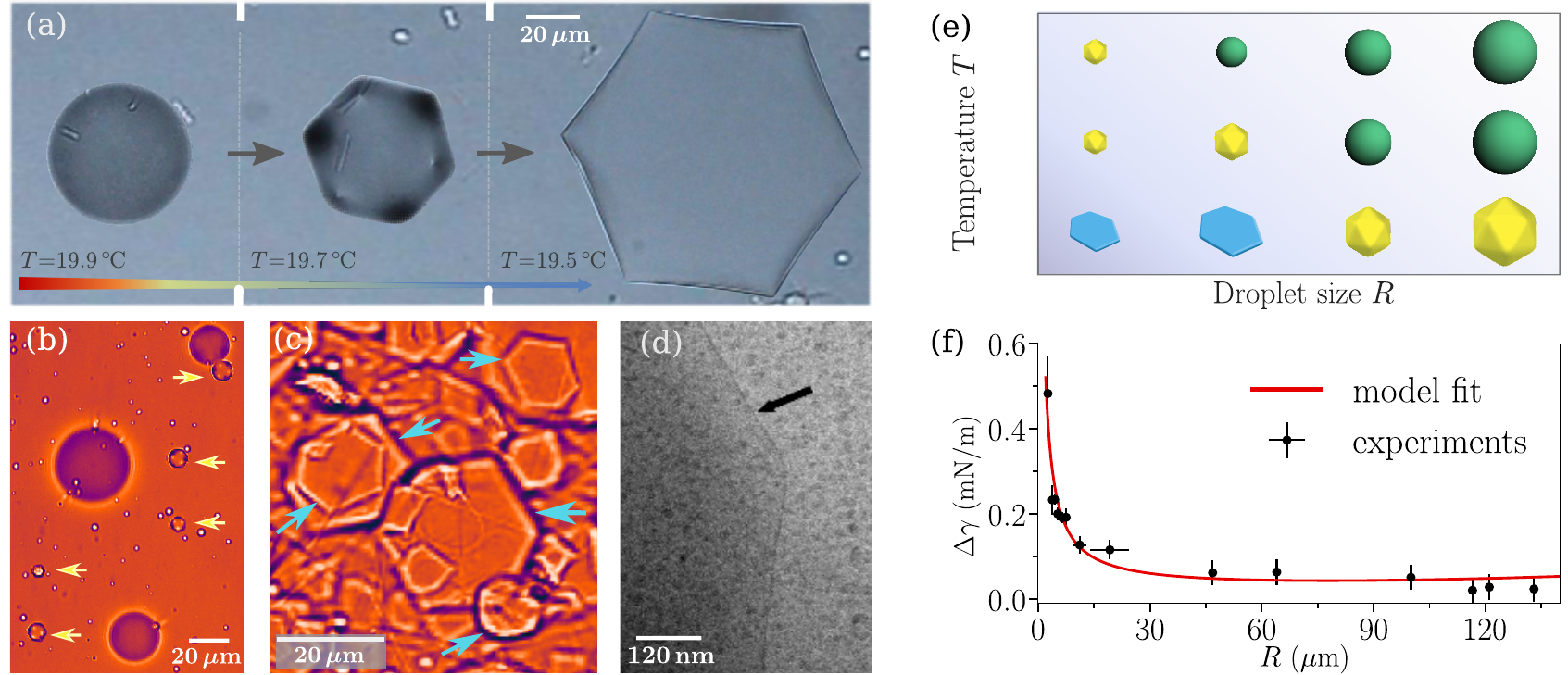}
    \caption{(a) Light microscopy snapshots of the volume-conserving faceting and flattening of a single spherical droplet as temperature is lowered. (b) Typical configuration of a polydispersed sample showing large spherical and small icosahedral droplets (highlighted by arrows). (c) Icosahedral droplets and platelets are oriented with a flat face orthogonal to the direction of gravity.   (d) CryoTEM image of the interface of an icosahedral droplet, revealing that the thickness of the crystalline structure is close to $3$ nm, corresponding solely to the interfacially-frozen monolayer (reproduced and modified with permission from \cite{Guttman2019}).  (e) Schematic of the phase diagram of the droplets based on the experimental observations, with the triangulated surfaces used in the numerical calculations of the energy [see \Eqref{eq:EDimFull}]. (f) Experimental estimate and theoretical fit of the difference  between the surface tension at the sphere-icosahedron and icosahedron-platelet transitions, $\Delta\gamma=\gamma_{\rm sph-ico}-\gamma_{\rm ico-pla}$, as a function of droplet radius \cite{Guttman2016-1}. See Supplemental Material \cite{SI} for details.}
    \label{fig:experimental}
\end{figure*}

Despite liquid drops representing the quintessential realization of spherical geometry across an extraordinary vast range of length scales, from stars down to micro and nanoscale aerosols, a variety of faceted polyhedral shapes has been reported in simple oil-in-water emulsions \cite{Sloutskin2005} and recently investigated by Denkov {\em et al}. \cite{Denkov2015} and Guttman {\em et al}. \cite{Guttman2016-1} ({\Fref{fig:experimental}a-c}). Unlike typical emulsion droplets, these are enclosed by an interfacially-frozen monolayer of alkane molecules and surfactants. Upon cooling, the droplets undergo a series of shape transformations: from spheres to icosahedra, to hexagonal platelets (\Fref{fig:experimental}a), to even more exotic shapes featuring tentacle-like protrusions \cite{Guttman2016-1}. The specific transition pathway is not universal, but depends on several factors, such as the oil and surfactant chemical composition, the cooling rate and the droplet size \cite{Guttman2016-1,Guttman2017,Denkov2015,Cholakova2016, Denkov2016}.

Whereas these fascinating experimental results are now reproducible (see Ref. \cite{Marin2020} for a recent overview), a convincing explanation of the physical mechanisms underpinning the sequence of shape transformations is still lacking, despite two alternative scenarios having been proposed \cite{Guttman2016-1,Guttman2016-2,Guttman2017, Guttman2019,Denkov2015,Denkov2016,Haas2017,Haas2019,Cholakova2016,Cholakova2019}.
The first, hereafter referred to as rotator phase mechanism \cite{Denkov2015, Haas2017}, revolves around the existence of a rotator phase in the proximity of the droplet's surface, whose estimated thickness ranges between $45$ \cite{Cholakova2019} and $300$ nm \cite{Denkov2015}, serving as a plastic scaffold for the observed shape transformations, especially across different flat morphologies \cite{Haas2017}. The second scenario, proposed in Refs. \cite{Guttman2016-1,Guttman2016-2} and referred to as elastic buckling mechanism, ascribes instead the transformations to a competition between the stretching elasticity of the frozen interfacial monolayer and surface tension. As the former consists of a triangular lattice \cite{Tamam2011} lying on a closed surface, its structure is geometrically frustrated and inevitably features topological defects, where the local sixfold rotational symmetry of the lattice is broken \cite{Nelson2002Defects}. Furthermore, as the elastic stress introduced by these defects can be relieved by increasing the local curvature \cite{BNT2000,Bowick2009}, such a crystalline monolayer is generally prone to buckling and faceting. The most prominent example of this mechanism is found in the context of viral capsids. In these, the trade-off between in-plane stresses, originating from the presence of twelve topologically required fivefold disclinations, and bending moments, resulting from out-of-plane deformations, drives the buckling of spheres into icosahedra \cite{Lidmar2003,Siber2006,Funkhouser2013}. The transition occurs at large values of the so called F{\"o}ppl-von K{\'a}rm{\'a}n number, $\young R^2/\bend$, expressing the ratio between the stretching and bending energy scale, with $\young$ the two-dimensional Young modulus, $R$ the system size (e.g. the capsid radius prior to buckling) and $\bend$ the bending rigidity. Thus, large viral capsids are energetically favored to be icosahedral, while small capsids are preferentially spherical.

Although potentially plausible to justify the observed sequence of shape transformations, neither of these scenarios succeeds in explaining all the experimental observations. Recent cryo-transmission electron (CryoTEM) micrographs revealed that the solid layer located at the oil-water interface of small icosahedral and other polyhedral droplets is only a few nanometers thick (\Fref{fig:experimental}d and Refs. \cite{Guttman2019,Tamam2011}), thus too thin to support the rotator phase mechanism, at least in the experimental set up pioneered by Guttman {\em et al.} \cite{Guttman2016-1,Guttman2016-2,Guttman2017,Guttman2019}. By contrast, the elastic buckling mechanism fails to reproduce the observed size dependence of the sphere-icosahedron-platelet transition. Denkov {\em et al.} noted that  while lowering the temperature, the smaller the droplet, the more the shape changes before reaching the bulk oil freezing point, with the largest droplets remaining spherical \cite{Denkov2015,Cholakova2016}. Consistently, by imaging several individual droplets upon slow cooling, Guttman {\em et al}. reported quantitatively that small droplets undergo faceting at higher temperatures compared to large droplets (see micrograph in {\Fref{fig:experimental}b} and the experimental data in  \Fref{fig:experimental}f, data from Ref. \cite{Guttman2016-1}, with slightly improved statistics). This cannot be explained either via the classic virus buckling picture, whose size-dependence is in fact opposite to that observed in icosahedral droplets \cite{Lidmar2003}, nor by postulating a similar interplay between defect-driven stretching elasticity and surface tension. As both the stretching and the surface energy scale like $R^{2}$, the latter implies that, depending on the ratio between $Y$ and the surface tension $\surfT$, at a given temperature, droplets are either always spherical (for small $Y/\surfT$ values) or always icosahedral (for large $Y/\surfT$ values), regardless of their size. Furthermore, in order for the elastic buckling mechanism to account for the icosahedra-platelets transition, one must assume $\bend\approx 10^{-1}\,k_{B}T$ \cite{Haas2019}, orders of magnitude smaller than the value $10^{3}\,k_{B}T$ estimated in Ref. \cite{Guttman2016-1}.

In this article, we resolve this dilemma by demonstrating that elastic buckling can, in fact, explain not only the sphere-icosahedron-platelet transition, but also the size dependence observed in experiments, provided it is augmented with the following mechanisms: the spontaneous curvature of the frozen alkane-surfactant monolayer and gravity. The starting point of our approach is the following functional describing the energy of a droplet of volume $V$ and density $\rho_{\rm oil}$ enclosed by a crystalline monolayer and suspended in water, namely:
\begin{equation}\label{eq:EFull}
E = \frac{1}{2} \int {\rm d}A\,\left[2 \surfT_0 + Y\sigma^{2} + 4\bend(H-H_{0})^{2} + g \Delta\rho z^{2}N_{z}\right]\;,
\end{equation}
where $\sigma=\sigma_{i}^{i}/Y$ is the trace of the covariant stress tensor $\sigma_{ij}$ arising in the interface in response to stretching deformations, $H$ is the droplet mean curvature taking $H>0$ for a sphere, $H_{0}$ the spontaneous curvature, $g$ the gravitational acceleration, $\Delta \rho = \rho_{\rm water}-\rho_{\rm oil}$ the oil-water density difference and $N_{z}$ the projection of the surface normal vector along the $z-$direction. The first term in Eq. \eqref{eq:EFull} corresponds to the standard surface energy. The second term accounts for the in-plane stretching originating from the combined effect of the Gaussian curvature and the topological defects. The third term describes the energetic cost of bending, in terms of the departure of the droplet mean curvature from its preferential value $H_{0}$. The latter arises in droplets as a consequence of the asymmetry of the adsorbed surfactant molecules \cite{Paunov2000}, and can be interpreted as a renormalization of the surface tension, analogous to that caused by an inhomogeneous Tolman length \cite{Wilhelmsen2015}:
\begin{equation}\label{eq:tolman}
\surfT' = \surfT - 4\bend H_{0} H\;,
\end{equation}
where $\surfT =\surfT_{0} + 2\bend H_{0}^{2}$ is the uniform part of the effective surface tension, inclusive of the contribution arising from the spontaneous curvature $H_{0}$. As we detail further below, this renormalization is instrumental to the observed size dependence. Since surface tension is the main restoring mechanism in spherical droplets, Eq. \eqref{eq:tolman} implies that for $H_0>0$, the smaller the droplet the less it is prone to return to a spherical shape once it is deformed. Finally, the last term in Eq. \eqref{eq:EFull} describes the gravity-buoyancy balance caused by the density difference $\Delta\rho$, here expressed as a surface integral by means of  the divergence theorem.

In the following, we assume that all the material parameters are uniform across the surface and we search for the lowest energy configuration of the droplet as a function of $\gamma$ and for various $\bend$, $H_0$, $\young$, and $\dens$ values. Consistently with experimental studies, $\gamma \sim T$ is the only material parameter strongly affected by temperature within the experimental range and can, therefore, be used as a proxy for temperature \cite{Guttman2016-2,Haas2017} (see also Ref. \cite{SI}). We compute the energy associated with three configurations depicted in Fig. \ref{fig:experimental}a by modeling surfaces via triangulated meshes with rounded vertices and edges \cite{SI}. The platelet, in particular, has a height-to-width ratio of around $1/10$  at the center, consistently with three-dimensional electron microscopy reconstructions \cite{Marin2019} (see also Fig. S1). Measuring energy in units of $\bend$ and length in units of $1/H_0$, we numerically calculate the dimensionless energy $\mathcal{E}=E/\bend$ for each surface as a function of the dimensionless size $r=H_{0}R$, with $R=[V/(4\pi/3)]^{1/3}$. Namely:
\begin{equation}
\label{eq:EDimFull}
\mathcal{E} = \mathcal{E}_{W}-\mathcal{E}_{H} r + (\dimSurf\mathcal{E}_{C}+\fvk\mathcal{E}_{S})r^{2}+\squash\mathcal{E}_{G}r^{4}\;,
\end{equation}
where the terms on the right-hand side denote the dimensionless form of the bending ($\mathcal{E}_{W}$, $\mathcal{E}_{H}$), capillary ($\mathcal{E}_{C}$), stretching ($\mathcal{E}_{S}$) and gravitational ($\mathcal{E}_{G}$) energies. The bending energy has been split into the so called Willmore functional, $\mathcal{E}_{W}= 2 \int {\rm d}A\,H^{2}$, and the contribution associated with the spontaneous curvature, $\mathcal{E}_{H}R=4 \int {\rm d}A\,H$. The numbers
\[
\fvk = \frac{\young}{\bend H_0^2}\;,\qquad
\dimSurf = \frac{\surfT}{\bend H_0^2}\;,\qquad
\squash = \frac{g\Delta\rho}{\bend H_0^4}\;,
\]
quantify the energetic cost of stretching, capillarity and gravity in comparison to bending, and they constitute the set of independent material parameters of the system. 

\begin{table}[t!]
	\begin{ruledtabular}
	\begin{tabular}[c]{cccc}
		 Energy & Sphere & Icosahedron & Platelet \\
		\hline
	    $\mathcal{E}_{W}$ & 25.4    & 49.2  & 209.6     \\
		$\mathcal{E}_{H}$ &  50.3    & 55.9  & 112.0      \\
		$\mathcal{E}_{C}$ & 12.6    & 12.9  & 34.6      \\	
		$\mathcal{E}_{S}$ & 0.0028  & 0.0013&  0.0016   \\
		$\mathcal{E}_{G}$ & 4.19    & 4.00  &  0.555     \\
	\end{tabular}
	\end{ruledtabular}
\caption{\label{tab:energy} Contributions to the dimensionless energy \Eqref{eq:EDimFull} for the spherical, icosahedral and hexagonal droplets. See Ref. \cite{SI} for details.}
\label{table:geometric}
\end{table}

\begin{figure}[t!]
    \centering
    \includegraphics[width=\columnwidth]{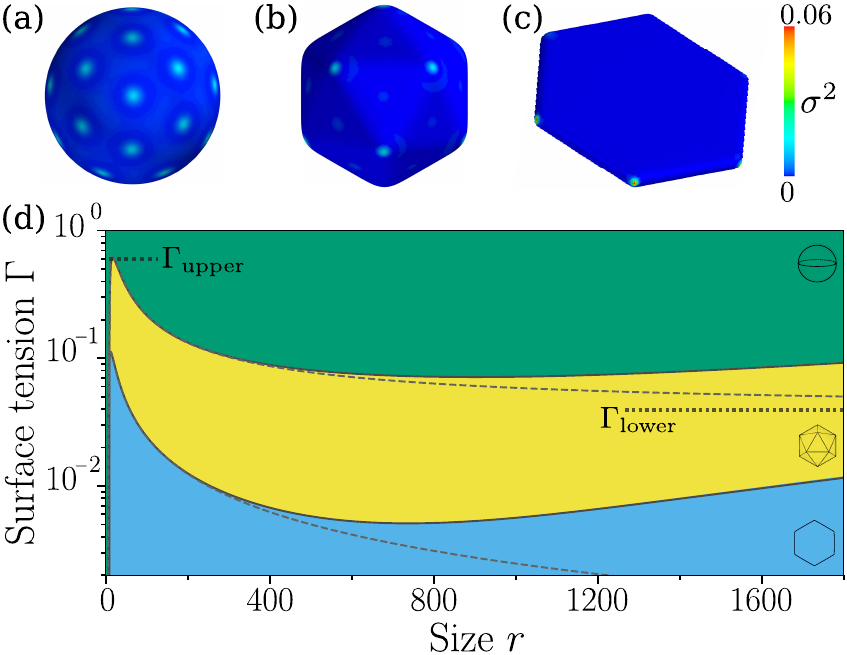}
    \caption{(a-c) Defect-induced stress field $\sigma^2$ on the crystalline interface with screening dislocations, computed from the charge density $\eta$ in \Eqref{eq:eta}.  (d) Morphological phase diagram in the $(r,\, \dimSurf)-$plane obtained by comparing the energies of spherical, icosahedral and platelet-shaped droplets calculated via Eq. \eqref{eq:EDimFull}.  
     The bound values of $\dimSurf$ for faceting, \Eqref{eq:lambda_conditions}, are indicated with dotted lines. The dashed lines correspond to the phase boundaries when buoyancy effects are ignored, while the solid ones are for $\squash=10^{-8}$. To create this diagram we used $\fvk=4$. 
     }
    \label{fig:phase_diagram}
\end{figure}

With the exception of $\mathcal{E}_{S}$, all energy contributions depend exclusively on the shape of the droplet and can be straightforwardly computed using our triangular meshes (Table \ref{table:geometric}).  In order to calculate the stretching energy, one needs to determine the local stress induced by the non-vanishing curvature of the droplet and by the topological defects populating the interfacial monolayer, by solving the Poisson equation $\nabla^{2}\sigma = \eta-K$, 
where $\nabla^{2}$ is the Laplace-Beltrami operator, $K$ the surface Gaussian curvature and $\eta$ is the topological charge density of the defect distribution \cite{Bowick2009}. In general, the latter consists of twelve topologically required, fivefold disclinations surrounded by a ``cloud'' of topologically neutral dislocations, which ease the distortion introduced by the disclinations, thereby reducing the elastic energy \cite{BNT2000,Bausch2003}. Following Ref. \cite{GarciaAguilar2020}, we express $\eta$ in terms of a discrete set of ``seed'' disclinations, coupled with a continuous distribution of screening dislocations. 
This yields:
\begin{equation}\label{eq:eta}
\eta 
= \sum_{\alpha=1}^{\mathcal{V}}\left(\frac{\pi}{3}\,q_{\alpha}-\Phi\right)\delta(\bm{r}-\bm{r}_{\alpha})
+ \frac{\Phi}{\mathcal{F}/\mathcal{V}}\sum_{\beta=1}^{\mathcal{F}}\delta(\bm{r}-\bm{r}_{\beta})\;,
\end{equation}
where $\mathcal{V}$ and $\mathcal{F}$ represent, respectively, the number of vertices and faces of the polyhedral droplets, $q_{\alpha}=6-z_{\alpha}$ is the topological charge of a $z_{\alpha}-$fold disclination located at position $\bm{r}_{\alpha}$ and $\Phi$ is the in-plane flux of the screening dislocations originating in proximity of the vertices and terminating at positions $\bm{r}_{\beta}$ in the bulk of the faces of the polyhedral droplets, where $K\approx 0$ (see Fig. S2 in \cite{SI}). This construction guarantees that $\int {\rm d}A\,\eta=4\pi$ such that, consistently with the divergence and the Gauss-Bonnet theorems, both sides of the stress equation vanish upon integration over the entire surface. The optimal dislocation flux $\Phi$ for a given surface is found by minimizing the energy $\mathcal{E}_{S}$ \cite{GarciaAguilar2020}, although all our results still hold qualitatively under the assumption of little to no screening (i.e. $\Phi=0$). 
Within this framework, we have numerically calculated and compared the  energies $\mathcal{E}_{\rm sph}$, $\mathcal{E}_{\rm ico}$ and $\mathcal{E}_{\rm pla}$ of the spherical, icosahedral and platelet conformations. The outcome of our analysis is summarized in Table \ref{tab:energy} and in the phase diagram of Fig. \ref{fig:phase_diagram}.

First, we focus on the faceting transition (i.e. sphere-icosahedron), for which buoyancy plays a marginal role (see $\mathcal{E}_{G}$ in Table \ref{tab:energy}) and thus can be temporarily neglected. In this case, the total energy is just a quadratic function of the dimensionless size $r$, from which the corresponding phase boundaries can be easily computed as shown by the dashed lines in Fig. \ref{fig:phase_diagram}. The defect configuration that minimizes the stretching energy  $\mathcal{E}_{S}$
consists of twelve fivefold disclinations (i.e. $q_{\alpha}=1$) approximately located at the vertices of an icosahedron and surrounded by  screening dislocations, so that $\eta\approx K$ in their vicinity. In the absence of restoring mechanisms favoring spherical shapes, as bending and capillarity, icosahedral shapes would be preferred over spherical ones, for \textit{any} temperature and droplet size. Since the icosahedron has a larger area and bending energy compared to a sphere of the same volume (both $\mathcal{E}_{W}$ and $\mathcal{E}_{H}$ diverge for a perfectly sharp icosahedron), these restoring mechanisms render the icosahedral conformation energetically optimal only at low temperature, where capillarity is sufficiently weak (i.e. the yellow region of Fig. \ref{fig:phase_diagram}). Furthermore, as surface tension becomes effectively smaller for decreasing droplet size due to the spontaneous curvature [\Eqref{eq:tolman}], smaller droplets generally undergo the faceting transition at higher temperature than large droplets, consistently with the experimental observations ({\Fref{fig:experimental}b} and Ref. \cite{Guttman2016-1}). At a fixed $\Gamma$ value, the critical size at which the transition takes place is found by solving the equation $\mathcal{E}_{\rm ico}=\mathcal{E}_{\rm sph}$ with respect to $r>0$. These solutions yield the range of parameters in which spherical and icosahedral droplets coexist, namely:
\begin{equation}
\label{eq:lambda_conditions}
0
< \dimSurf + \fvk\,\frac{\Delta\mathcal{E}_{S}}{\Delta\mathcal{E}_{C}} 
< \frac{\Delta\mathcal{E}_{H}^2}{4\Delta\mathcal{E}_{W}\Delta\mathcal{E}_{C}}\;,
\end{equation}
where $\Delta\mathcal{E}_{i}$ ($i=S,\,C,\,H,\,W$) labels the various contribution of the \ $\mathcal{E}_{\rm ico}-\mathcal{E}_{\rm sph}$ energy difference. The upper bound of this inequality corresponds to the peak of the yellow region in Fig. \ref{fig:phase_diagram}d, above which capillarity dominates and droplets are always spherical regardless of their size. The abscissa of the peak of the sphere-icosahedron phase boundary, $r= 2\Delta\mathcal{E}_{W}/\Delta\mathcal{E}_{H}$, approximates the minimal droplet radius $r_{\min}$ at which icosahedral droplets can be found. For $r<r_{\min}$, the Gaussian curvature of the sphere is sufficiently large to accommodate the angular deficit introduced by the fivefold disclinations and faceting does not occur, except in the limit of vanishing bending rigidity. The lower limit $\dimSurf=-\fvk \Delta\mathcal{E}_{S}/\Delta \mathcal{E}_{C}$ in \Eqref{eq:lambda_conditions}, defines a lower critical surface tension, indicated in Fig. \ref{fig:phase_diagram},  below which all droplets of size $r>r_{\min}$  are icosahedral. Finally, buoyancy only affects the large$-r$ region of the phase diagram by favoring icosahedra over spheres (color boundary in \Fref{fig:phase_diagram}d). In general, the gravitational energy can be lowered by reducing the distance between the droplet center of mass and the top wall of the sample container. In particular, for regular polyhedra with ${\rm D}_{nh}$ and ${\rm D}_{nd}$ point group, $\mathcal{E}_{G}R = (4\pi/3)\,h/2$ with $h$ the height of the droplet, and this can be achieved by aligning one of the flat faces orthogonally to the $z-$direction. As $h/2<R$ in the faceted droplets of conserved volume, buoyancy widens the icosahedral phase at large $r$ values. 

\begin{figure}[t!]
    \centering
    \includegraphics[width=\columnwidth]{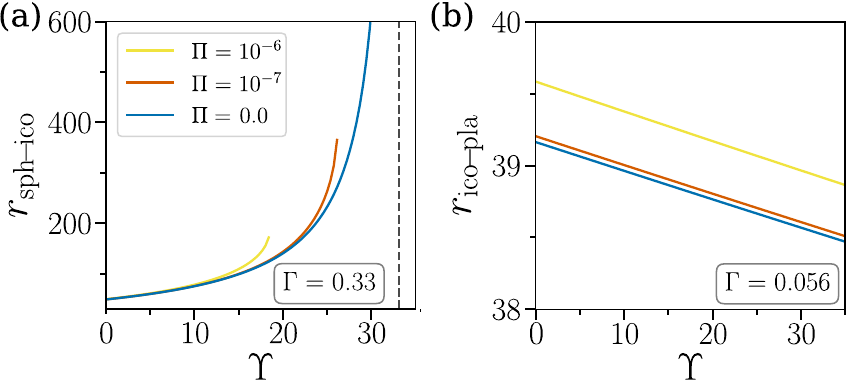}
    \caption{Critical size for the faceting (a) and flattening (b) transitions, obtained by solving the equations $\mathcal{E}_{\rm sph}=\mathcal{E}_{\rm ico}$ and $\mathcal{E}_{\rm ico}=\mathcal{E}_{\rm pla}$ with respect to $r$, versus the $\fvk$ parameter. Note in (a) that for each $\squash$, there is a maximal $\fvk$ beyond which no critical radius exist for faceting, as all droplets are icosahedral. 
    }
    \label{fig:material_parameters}
\end{figure}

Next we focus on the flattening transition (i.e. icosahedron-platelet). Experimentally, this is observed at ultra-low (or even transiently negative) values of the surface tension \cite{Guttman2016-2}. In this regime, the system lacks of the main restoring mechanism favoring spherical shapes and the icosahedral configuration represents the absolute minimum of the elasto-capillary energy for large $\fvk$ values. 
As the area of a platelet is larger than that of an icosahedron with the same volume, in the absence of spontaneous curvature and/or gravitational effects, the  droplet flattening can only occur at negative $\dimSurf$ values. A positive spontaneous curvature, by contrast, favors platelets over icosahedra at small $r$ values as a consequence of the larger mean curvature at the edges and vertices. 
Furthermore, upon orienting orthogonally with respect to the $z-$direction, platelets can raise their center of mass to reduce their gravitational energy to arbitrarily small values, at the cost of increasing the stretching energy by pairwise merging the twelve fivefold disclinations into six fourfold disclinations (i.e. with charge $q_{\alpha}=2$). 
This causes icosahedral droplets to morph into hexagonal platelets, where the Gaussian curvature at the vertices is sufficiently large to compensate for the elastic distortion introduced by the higher topological charge, resulting into a modest increase of the stretching energy (see Table \ref{tab:energy}). Because of spontaneous curvature at smaller scales and buoyancy at larger ones, flattening is possible at very low but still positive surface tension values (see \Fref{fig:phase_diagram}), consistently with experimental observations. 

To assess the significance of our predictions, we have fitted the difference $\Delta\gamma=\gamma_{\rm sph-ico}-\gamma_{\rm ico-pla}$ between the surface tension at the sphere-icosahedron and icosahedron-platelet transitions as a function of the droplet radius, upon fixing $\Delta\rho=0.25\,{\rm g}/{\rm cm}^{3}$ \cite{Guttman2017,Haas2019} and $\kappa= 10^{3}\,k_{B}T$ \cite{Guttman2016-1} (see red line in \Fref{fig:experimental}f). From the fit we obtain $\young\approx 4.4\,{\rm mN}/{\rm m}$ and $H_0^{-1}\approx 58\,{\rm nm}$, consistently with current knowledge (see Ref. \cite{SI} for details).

In conclusion, we look at the effect of the mechanical properties of the frozen interfacial monolayer, embodied in the number $\fvk$, for both transitions. Assuming $\dimSurf$ fixed, we solve the equations $\mathcal{E}_{\rm sph}=\mathcal{E}_{\rm ico}$ and $\mathcal{E}_{\rm ico}=\mathcal{E}_{\rm pla}$ with respect to the $r$ and $\fvk$ in the proximity of the sphere-icosahedron and icosahedron-platelet phase boundaries.  The corresponding solutions are displayed in Fig. \ref{fig:material_parameters} for $\dimSurf=0.33$ and $\dimSurf=0.056$. 
These are intermediate values from the range in \Eqref{eq:lambda_conditions} when $\fvk=4$, where we observe the coexistence of different shapes for differently sized droplets, as in experiments. The critical droplet size associated with the faceting transition increases with $\fvk$ until a limiting value where all droplets are icosahedral at any $r>r_{\min}$. For $\squash=0$, this corresponds to the dashed vertical line in {\Fref{fig:material_parameters}a}. 
By contrast, the critical radius associated with the flattening transition ({\Fref{fig:material_parameters}b}), is a slowly decreasing function of $\fvk$, since only at smaller sizes is the curvature gain enough to compensate for the slight increase in stretching. 
As $\fvk \rightarrow 0$, however, the two critical radii are comparable in magnitude, indicating that decreasing $\Upsilon$ narrows the gap between the faceting and flattening transition, and also allows for the possible coexistence of spheres and platelets at the same temperature by reducing the range of the icosahedral phase. Finally, both critical radii increase under the effect of gravity, as this facilitates departure from isotropic shapes.
In panel (a), the limiting $\fvk$ value where droplets of all sizes are icosahedral, decreases for increasing $\squash$ values. 

In summary, we theoretically addressed the mechanical origin of the faceting and flattening transitions that, starting from the pioneering work of Denkov \cite{Denkov2015} and Guttman {\em et al}. \cite{Guttman2016-1} have been systematically reported in emulsion droplets stabilized by a frozen layer of alkane molecules and surfactants. Using a combination of continuum mechanics and three-dimensional computer modeling, we demonstrated that both transitions originate from the fourfold interplay between defect-driven stretching, bending elasticity, capillarity and gravity. In particular, we highlighted the importance of the spontaneous curvature of the interface monolayer, that, by effectively hindering the magnitude of capillary forces, allows one to reproduce the counterintuitive size-dependence observed in the experiments. While more advanced theoretical models may be needed to describe the behavior of multicomponent droplets \cite{Marin2020}, our present work provides the basis for these more complex future models.

\acknowledgements

We thank Ray Goldstein and Moshe Deutsch for illuminating discussions and Shir R. Liber and Alexander V. Butenko for technical assistance. This work is partially supported by Netherlands Organisation for Scientific Research (NWO/OCW), as part of the D-ITP program (I.G.A. and L.G.), the Vidi scheme (P.F. and L.G.), the Frontiers of Nanoscience program (L.G.) and the Israel Science Foundation, grant no. 1779/17 (E. S.). 

\nocite{Sullivan2008,SurfaceEvolver,AshcroftMermim,Yefet2014-2,Barker1963}
\bibliography{bibDroplets}{}

\begin{thebibliography}{30}%
\makeatletter
\providecommand \@ifxundefined [1]{%
 \@ifx{#1\undefined}
}%
\providecommand \@ifnum [1]{%
 \ifnum #1\expandafter \@firstoftwo
 \else \expandafter \@secondoftwo
 \fi
}%
\providecommand \@ifx [1]{%
 \ifx #1\expandafter \@firstoftwo
 \else \expandafter \@secondoftwo
 \fi
}%
\providecommand \natexlab [1]{#1}%
\providecommand \enquote  [1]{``#1''}%
\providecommand \bibnamefont  [1]{#1}%
\providecommand \bibfnamefont [1]{#1}%
\providecommand \citenamefont [1]{#1}%
\providecommand \href@noop [0]{\@secondoftwo}%
\providecommand \href [0]{\begingroup \@sanitize@url \@href}%
\providecommand \@href[1]{\@@startlink{#1}\@@href}%
\providecommand \@@href[1]{\endgroup#1\@@endlink}%
\providecommand \@sanitize@url [0]{\catcode `\\12\catcode `\$12\catcode
  `\&12\catcode `\#12\catcode `\^12\catcode `\_12\catcode `\%12\relax}%
\providecommand \@@startlink[1]{}%
\providecommand \@@endlink[0]{}%
\providecommand \url  [0]{\begingroup\@sanitize@url \@url }%
\providecommand \@url [1]{\endgroup\@href {#1}{\urlprefix }}%
\providecommand \urlprefix  [0]{URL }%
\providecommand \Eprint [0]{\href }%
\providecommand \doibase [0]{http://dx.doi.org/}%
\providecommand \selectlanguage [0]{\@gobble}%
\providecommand \bibinfo  [0]{\@secondoftwo}%
\providecommand \bibfield  [0]{\@secondoftwo}%
\providecommand \translation [1]{[#1]}%
\providecommand \BibitemOpen [0]{}%
\providecommand \bibitemStop [0]{}%
\providecommand \bibitemNoStop [0]{.\EOS\space}%
\providecommand \EOS [0]{\spacefactor3000\relax}%
\providecommand \BibitemShut  [1]{\csname bibitem#1\endcsname}%
\let\auto@bib@innerbib\@empty
\bibitem [{\citenamefont {Sloutskin}\ \emph {et~al.}(2005)\citenamefont
  {Sloutskin}, \citenamefont {Bain}, \citenamefont {Ocko},\ and\ \citenamefont
  {Deutsch}}]{Sloutskin2005}%
  \BibitemOpen
  \bibfield  {author} {\bibinfo {author} {\bibfnamefont {E.}~\bibnamefont
  {Sloutskin}}, \bibinfo {author} {\bibfnamefont {C.~D.}\ \bibnamefont {Bain}},
  \bibinfo {author} {\bibfnamefont {B.~M.}\ \bibnamefont {Ocko}}, \ and\
  \bibinfo {author} {\bibfnamefont {M.}~\bibnamefont {Deutsch}},\ }\href
  {\doibase 10.1039/B405969G} {\bibfield  {journal} {\bibinfo  {journal}
  {Faraday Discuss.}\ }\textbf {\bibinfo {volume} {129}},\ \bibinfo {pages}
  {339} (\bibinfo {year} {2005})},\ \bibinfo {note} {see in particular, the
  General Discussion, p. 353-366}\BibitemShut {NoStop}%
\bibitem [{\citenamefont {Denkov}\ \emph {et~al.}(2015)\citenamefont {Denkov},
  \citenamefont {Tcholakova}, \citenamefont {Lesov}, \citenamefont
  {Cholakova},\ and\ \citenamefont {Smoukov}}]{Denkov2015}%
  \BibitemOpen
  \bibfield  {author} {\bibinfo {author} {\bibfnamefont {N.}~\bibnamefont
  {Denkov}}, \bibinfo {author} {\bibfnamefont {S.}~\bibnamefont {Tcholakova}},
  \bibinfo {author} {\bibfnamefont {I.}~\bibnamefont {Lesov}}, \bibinfo
  {author} {\bibfnamefont {D.}~\bibnamefont {Cholakova}}, \ and\ \bibinfo
  {author} {\bibfnamefont {S.~K.}\ \bibnamefont {Smoukov}},\ }\href {\doibase
  10.1038/nature16189} {\bibfield  {journal} {\bibinfo  {journal} {Nature}\
  }\textbf {\bibinfo {volume} {528}},\ \bibinfo {pages} {392–395} (\bibinfo
  {year} {2015})}\BibitemShut {NoStop}%
\bibitem [{\citenamefont {Guttman}\ \emph
  {et~al.}(2016{\natexlab{a}})\citenamefont {Guttman}, \citenamefont {Sapir},
  \citenamefont {Schultz}, \citenamefont {Butenko}, \citenamefont {Ocko},
  \citenamefont {Deutsch},\ and\ \citenamefont {Sloutskin}}]{Guttman2016-1}%
  \BibitemOpen
  \bibfield  {author} {\bibinfo {author} {\bibfnamefont {S.}~\bibnamefont
  {Guttman}}, \bibinfo {author} {\bibfnamefont {Z.}~\bibnamefont {Sapir}},
  \bibinfo {author} {\bibfnamefont {M.}~\bibnamefont {Schultz}}, \bibinfo
  {author} {\bibfnamefont {A.~V.}\ \bibnamefont {Butenko}}, \bibinfo {author}
  {\bibfnamefont {B.~M.}\ \bibnamefont {Ocko}}, \bibinfo {author}
  {\bibfnamefont {M.}~\bibnamefont {Deutsch}}, \ and\ \bibinfo {author}
  {\bibfnamefont {E.}~\bibnamefont {Sloutskin}},\ }\href {\doibase
  10.1073/pnas.1515614113} {\bibfield  {journal} {\bibinfo  {journal} {Proc.
  Natl. Acad. Sci. U. S. A.}\ }\textbf {\bibinfo {volume} {113}},\ \bibinfo
  {pages} {493} (\bibinfo {year} {2016}{\natexlab{a}})}\BibitemShut {NoStop}%
\bibitem [{\citenamefont {Guttman}\ \emph {et~al.}(2017)\citenamefont
  {Guttman}, \citenamefont {Sapir}, \citenamefont {Ocko}, \citenamefont
  {Deutsch},\ and\ \citenamefont {Sloutskin}}]{Guttman2017}%
  \BibitemOpen
  \bibfield  {author} {\bibinfo {author} {\bibfnamefont {S.}~\bibnamefont
  {Guttman}}, \bibinfo {author} {\bibfnamefont {Z.}~\bibnamefont {Sapir}},
  \bibinfo {author} {\bibfnamefont {B.~M.}\ \bibnamefont {Ocko}}, \bibinfo
  {author} {\bibfnamefont {M.}~\bibnamefont {Deutsch}}, \ and\ \bibinfo
  {author} {\bibfnamefont {E.}~\bibnamefont {Sloutskin}},\ }\href {\doibase
  10.1021/acs.langmuir.6b02926} {\bibfield  {journal} {\bibinfo  {journal}
  {Langmuir}\ }\textbf {\bibinfo {volume} {33}},\ \bibinfo {pages}
  {1305–1314} (\bibinfo {year} {2017})}\BibitemShut {NoStop}%
\bibitem [{\citenamefont {Cholakova}\ \emph {et~al.}(2016)\citenamefont
  {Cholakova}, \citenamefont {Denkov}, \citenamefont {Tcholakova},
  \citenamefont {Lesov},\ and\ \citenamefont {Smoukov}}]{Cholakova2016}%
  \BibitemOpen
  \bibfield  {author} {\bibinfo {author} {\bibfnamefont {D.}~\bibnamefont
  {Cholakova}}, \bibinfo {author} {\bibfnamefont {N.}~\bibnamefont {Denkov}},
  \bibinfo {author} {\bibfnamefont {S.}~\bibnamefont {Tcholakova}}, \bibinfo
  {author} {\bibfnamefont {I.}~\bibnamefont {Lesov}}, \ and\ \bibinfo {author}
  {\bibfnamefont {S.~K.}\ \bibnamefont {Smoukov}},\ }\href {\doibase
  10.1016/j.cis.2016.06.002} {\bibfield  {journal} {\bibinfo  {journal} {Adv.
  Colloid Interface Sci.}\ }\textbf {\bibinfo {volume} {235}},\ \bibinfo
  {pages} {90–107} (\bibinfo {year} {2016})}\BibitemShut {NoStop}%
\bibitem [{\citenamefont {Denkov}\ \emph {et~al.}(2016)\citenamefont {Denkov},
  \citenamefont {Cholakova}, \citenamefont {Tcholakova},\ and\ \citenamefont
  {Smoukov}}]{Denkov2016}%
  \BibitemOpen
  \bibfield  {author} {\bibinfo {author} {\bibfnamefont {N.}~\bibnamefont
  {Denkov}}, \bibinfo {author} {\bibfnamefont {D.}~\bibnamefont {Cholakova}},
  \bibinfo {author} {\bibfnamefont {S.}~\bibnamefont {Tcholakova}}, \ and\
  \bibinfo {author} {\bibfnamefont {S.~K.}\ \bibnamefont {Smoukov}},\ }\href
  {\doibase 10.1021/acs.langmuir.6b01626} {\bibfield  {journal} {\bibinfo
  {journal} {Langmuir}\ }\textbf {\bibinfo {volume} {32}},\ \bibinfo {pages}
  {7985–7991} (\bibinfo {year} {2016})}\BibitemShut {NoStop}%
\bibitem [{\citenamefont {Marin}\ \emph {et~al.}(2020)\citenamefont {Marin},
  \citenamefont {Tkachev}, \citenamefont {Sloutskin},\ and\ \citenamefont
  {Deutsch}}]{Marin2020}%
  \BibitemOpen
  \bibfield  {author} {\bibinfo {author} {\bibfnamefont {O.}~\bibnamefont
  {Marin}}, \bibinfo {author} {\bibfnamefont {M.}~\bibnamefont {Tkachev}},
  \bibinfo {author} {\bibfnamefont {E.}~\bibnamefont {Sloutskin}}, \ and\
  \bibinfo {author} {\bibfnamefont {M.}~\bibnamefont {Deutsch}},\ }\href
  {\doibase 10.1016/j.cocis.2020.05.006} {\bibfield  {journal} {\bibinfo
  {journal} {Curr. Opin. Colloid Interface Sci.}\ }\textbf {\bibinfo {volume}
  {49}},\ \bibinfo {pages} {107} (\bibinfo {year} {2020})}\BibitemShut
  {NoStop}%
\bibitem [{\citenamefont {Guttman}\ \emph
  {et~al.}(2016{\natexlab{b}})\citenamefont {Guttman}, \citenamefont {Ocko},
  \citenamefont {Deutsch},\ and\ \citenamefont {Sloutskin}}]{Guttman2016-2}%
  \BibitemOpen
  \bibfield  {author} {\bibinfo {author} {\bibfnamefont {S.}~\bibnamefont
  {Guttman}}, \bibinfo {author} {\bibfnamefont {B.~M.}\ \bibnamefont {Ocko}},
  \bibinfo {author} {\bibfnamefont {M.}~\bibnamefont {Deutsch}}, \ and\
  \bibinfo {author} {\bibfnamefont {E.}~\bibnamefont {Sloutskin}},\ }\href
  {\doibase 10.1016/j.cocis.2016.02.002} {\bibfield  {journal} {\bibinfo
  {journal} {Curr. Opin. Colloid Interface Sci.}\ }\textbf {\bibinfo {volume}
  {22}},\ \bibinfo {pages} {35–40} (\bibinfo {year}
  {2016}{\natexlab{b}})}\BibitemShut {NoStop}%
\bibitem [{\citenamefont {Guttman}\ \emph {et~al.}(2019)\citenamefont
  {Guttman}, \citenamefont {Kesselman}, \citenamefont {Jacob}, \citenamefont
  {Marin}, \citenamefont {Danino}, \citenamefont {Deutsch},\ and\ \citenamefont
  {Sloutskin}}]{Guttman2019}%
  \BibitemOpen
  \bibfield  {author} {\bibinfo {author} {\bibfnamefont {S.}~\bibnamefont
  {Guttman}}, \bibinfo {author} {\bibfnamefont {E.}~\bibnamefont {Kesselman}},
  \bibinfo {author} {\bibfnamefont {A.}~\bibnamefont {Jacob}}, \bibinfo
  {author} {\bibfnamefont {O.}~\bibnamefont {Marin}}, \bibinfo {author}
  {\bibfnamefont {D.}~\bibnamefont {Danino}}, \bibinfo {author} {\bibfnamefont
  {M.}~\bibnamefont {Deutsch}}, \ and\ \bibinfo {author} {\bibfnamefont
  {E.}~\bibnamefont {Sloutskin}},\ }\href {\doibase
  10.1021/acs.nanolett.9b00594} {\bibfield  {journal} {\bibinfo  {journal}
  {Nano Lett.}\ }\textbf {\bibinfo {volume} {19}},\ \bibinfo {pages} {3161}
  (\bibinfo {year} {2019})}\BibitemShut {NoStop}%
\bibitem [{\citenamefont {Haas}\ \emph {et~al.}(2017)\citenamefont {Haas},
  \citenamefont {Goldstein}, \citenamefont {Smoukov}, \citenamefont
  {Cholakova},\ and\ \citenamefont {Denkov}}]{Haas2017}%
  \BibitemOpen
  \bibfield  {author} {\bibinfo {author} {\bibfnamefont {P.~A.}\ \bibnamefont
  {Haas}}, \bibinfo {author} {\bibfnamefont {R.~E.}\ \bibnamefont {Goldstein}},
  \bibinfo {author} {\bibfnamefont {S.~K.}\ \bibnamefont {Smoukov}}, \bibinfo
  {author} {\bibfnamefont {D.}~\bibnamefont {Cholakova}}, \ and\ \bibinfo
  {author} {\bibfnamefont {N.}~\bibnamefont {Denkov}},\ }\href {\doibase
  10.1103/PhysRevLett.118.088001} {\bibfield  {journal} {\bibinfo  {journal}
  {Phys. Rev. Lett.}\ }\textbf {\bibinfo {volume} {118}},\ \bibinfo {pages}
  {088001} (\bibinfo {year} {2017})}\BibitemShut {NoStop}%
\bibitem [{\citenamefont {Haas}\ \emph {et~al.}(2019)\citenamefont {Haas},
  \citenamefont {Cholakova}, \citenamefont {Denkov}, \citenamefont
  {Goldstein},\ and\ \citenamefont {Smoukov}}]{Haas2019}%
  \BibitemOpen
  \bibfield  {author} {\bibinfo {author} {\bibfnamefont {P.~A.}\ \bibnamefont
  {Haas}}, \bibinfo {author} {\bibfnamefont {D.}~\bibnamefont {Cholakova}},
  \bibinfo {author} {\bibfnamefont {N.}~\bibnamefont {Denkov}}, \bibinfo
  {author} {\bibfnamefont {R.~E.}\ \bibnamefont {Goldstein}}, \ and\ \bibinfo
  {author} {\bibfnamefont {S.~K.}\ \bibnamefont {Smoukov}},\ }\href {\doibase
  10.1103/PhysRevResearch.1.023017} {\bibfield  {journal} {\bibinfo  {journal}
  {Phys. Rev. Research}\ }\textbf {\bibinfo {volume} {1}},\ \bibinfo {pages}
  {023017} (\bibinfo {year} {2019})}\BibitemShut {NoStop}%
\bibitem [{\citenamefont {Cholakova}\ \emph {et~al.}(2019)\citenamefont
  {Cholakova}, \citenamefont {Denkov}, \citenamefont {Tcholakova},
  \citenamefont {Valkova},\ and\ \citenamefont {Smoukov}}]{Cholakova2019}%
  \BibitemOpen
  \bibfield  {author} {\bibinfo {author} {\bibfnamefont {D.}~\bibnamefont
  {Cholakova}}, \bibinfo {author} {\bibfnamefont {N.}~\bibnamefont {Denkov}},
  \bibinfo {author} {\bibfnamefont {S.}~\bibnamefont {Tcholakova}}, \bibinfo
  {author} {\bibfnamefont {Z.}~\bibnamefont {Valkova}}, \ and\ \bibinfo
  {author} {\bibfnamefont {S.~K.}\ \bibnamefont {Smoukov}},\ }\href {\doibase
  10.1021/acs.langmuir.8b02771} {\bibfield  {journal} {\bibinfo  {journal}
  {Langmuir}\ }\textbf {\bibinfo {volume} {35}},\ \bibinfo {pages} {5484}
  (\bibinfo {year} {2019})}\BibitemShut {NoStop}%
\bibitem [{\citenamefont {Tamam}\ \emph {et~al.}(2011)\citenamefont {Tamam},
  \citenamefont {Pontoni}, \citenamefont {Sapir}, \citenamefont {Yefet},
  \citenamefont {Sloutskin}, \citenamefont {Ocko}, \citenamefont {Reichert},\
  and\ \citenamefont {Deutsch}}]{Tamam2011}%
  \BibitemOpen
  \bibfield  {author} {\bibinfo {author} {\bibfnamefont {L.}~\bibnamefont
  {Tamam}}, \bibinfo {author} {\bibfnamefont {D.}~\bibnamefont {Pontoni}},
  \bibinfo {author} {\bibfnamefont {Z.}~\bibnamefont {Sapir}}, \bibinfo
  {author} {\bibfnamefont {S.}~\bibnamefont {Yefet}}, \bibinfo {author}
  {\bibfnamefont {E.}~\bibnamefont {Sloutskin}}, \bibinfo {author}
  {\bibfnamefont {B.~M.}\ \bibnamefont {Ocko}}, \bibinfo {author}
  {\bibfnamefont {H.}~\bibnamefont {Reichert}}, \ and\ \bibinfo {author}
  {\bibfnamefont {M.}~\bibnamefont {Deutsch}},\ }\href {\doibase
  10.1073/pnas.1014100108} {\bibfield  {journal} {\bibinfo  {journal} {Proc.
  Natl Acad. Sci. U. S. A.}\ }\textbf {\bibinfo {volume} {108}},\ \bibinfo
  {pages} {5522} (\bibinfo {year} {2011})}\BibitemShut {NoStop}%
\bibitem [{\citenamefont {Nelson}(2002)}]{Nelson2002Defects}%
  \BibitemOpen
  \bibfield  {author} {\bibinfo {author} {\bibfnamefont {D.~R.}\ \bibnamefont
  {Nelson}},\ }\href@noop {} {\emph {\bibinfo {title} {Defects and Geometry in
  Condensed Matter Physics}}}\ (\bibinfo  {publisher} {Cambridge University
  Press},\ \bibinfo {year} {2002})\BibitemShut {NoStop}%
\bibitem [{\citenamefont {Bowick}\ \emph {et~al.}(2000)\citenamefont {Bowick},
  \citenamefont {Nelson},\ and\ \citenamefont {Travesset}}]{BNT2000}%
  \BibitemOpen
  \bibfield  {author} {\bibinfo {author} {\bibfnamefont {M.~J.}\ \bibnamefont
  {Bowick}}, \bibinfo {author} {\bibfnamefont {D.~R.}\ \bibnamefont {Nelson}},
  \ and\ \bibinfo {author} {\bibfnamefont {A.}~\bibnamefont {Travesset}},\
  }\href {\doibase 10.1103/PhysRevB.62.8738} {\bibfield  {journal} {\bibinfo
  {journal} {Phys. Rev. B}\ }\textbf {\bibinfo {volume} {62}},\ \bibinfo
  {pages} {8738} (\bibinfo {year} {2000})}\BibitemShut {NoStop}%
\bibitem [{\citenamefont {Bowick}\ and\ \citenamefont
  {Giomi}(2009)}]{Bowick2009}%
  \BibitemOpen
  \bibfield  {author} {\bibinfo {author} {\bibfnamefont {M.~J.}\ \bibnamefont
  {Bowick}}\ and\ \bibinfo {author} {\bibfnamefont {L.}~\bibnamefont {Giomi}},\
  }\href {\doibase 10.1080/00018730903043166} {\bibfield  {journal} {\bibinfo
  {journal} {Adv. Phys.}\ }\textbf {\bibinfo {volume} {58}},\ \bibinfo {pages}
  {449} (\bibinfo {year} {2009})}\BibitemShut {NoStop}%
\bibitem [{\citenamefont {Lidmar}\ \emph {et~al.}(2003)\citenamefont {Lidmar},
  \citenamefont {Mirny},\ and\ \citenamefont {Nelson}}]{Lidmar2003}%
  \BibitemOpen
  \bibfield  {author} {\bibinfo {author} {\bibfnamefont {J.}~\bibnamefont
  {Lidmar}}, \bibinfo {author} {\bibfnamefont {L.}~\bibnamefont {Mirny}}, \
  and\ \bibinfo {author} {\bibfnamefont {D.~R.}\ \bibnamefont {Nelson}},\
  }\href {\doibase 10.1103/PhysRevE.68.051910} {\bibfield  {journal} {\bibinfo
  {journal} {Phys. Rev. E}\ }\textbf {\bibinfo {volume} {68}},\ \bibinfo
  {pages} {051910} (\bibinfo {year} {2003})}\BibitemShut {NoStop}%
\bibitem [{\citenamefont {\ifmmode~\check{S}\else
  \v{S}\fi{}iber}(2006)}]{Siber2006}%
  \BibitemOpen
  \bibfield  {author} {\bibinfo {author} {\bibfnamefont {A.}~\bibnamefont
  {\ifmmode~\check{S}\else \v{S}\fi{}iber}},\ }\href {\doibase
  10.1103/PhysRevE.73.061915} {\bibfield  {journal} {\bibinfo  {journal} {Phys.
  Rev. E}\ }\textbf {\bibinfo {volume} {73}},\ \bibinfo {pages} {061915}
  (\bibinfo {year} {2006})}\BibitemShut {NoStop}%
\bibitem [{\citenamefont {Funkhouser}\ \emph {et~al.}(2013)\citenamefont
  {Funkhouser}, \citenamefont {Sknepnek},\ and\ \citenamefont {{Olvera de la
  Cruz}}}]{Funkhouser2013}%
  \BibitemOpen
  \bibfield  {author} {\bibinfo {author} {\bibfnamefont {C.~M.}\ \bibnamefont
  {Funkhouser}}, \bibinfo {author} {\bibfnamefont {R.}~\bibnamefont
  {Sknepnek}}, \ and\ \bibinfo {author} {\bibfnamefont {M.}~\bibnamefont
  {{Olvera de la Cruz}}},\ }\href {\doibase 10.1039/C2SM26607E} {\bibfield
  {journal} {\bibinfo  {journal} {Soft Matter}\ }\textbf {\bibinfo {volume}
  {9}},\ \bibinfo {pages} {60–68} (\bibinfo {year} {2013})}\BibitemShut
  {NoStop}%
\bibitem [{SI()}]{SI}%
  \BibitemOpen
  \href@noop {} {}\bibinfo {note} {See Supplemental Material at [URL will be
  inserted by publisher] for details on the numerical computation of
  \Eqref{eq:EDimFull} for the different droplet shapes}\BibitemShut {NoStop}%
\bibitem [{\citenamefont {Paunov}\ \emph {et~al.}(2000)\citenamefont {Paunov},
  \citenamefont {Sandler},\ and\ \citenamefont {Kaler}}]{Paunov2000}%
  \BibitemOpen
  \bibfield  {author} {\bibinfo {author} {\bibfnamefont {V.~N.}\ \bibnamefont
  {Paunov}}, \bibinfo {author} {\bibfnamefont {S.~I.}\ \bibnamefont {Sandler}},
  \ and\ \bibinfo {author} {\bibfnamefont {E.~W.}\ \bibnamefont {Kaler}},\
  }\href {\doibase 10.1021/la000367h} {\bibfield  {journal} {\bibinfo
  {journal} {Langmuir}\ }\textbf {\bibinfo {volume} {16}},\ \bibinfo {pages}
  {8917} (\bibinfo {year} {2000})}\BibitemShut {NoStop}%
\bibitem [{\citenamefont {Wilhelmsen}\ \emph {et~al.}(2015)\citenamefont
  {Wilhelmsen}, \citenamefont {Bedeaux},\ and\ \citenamefont
  {Reguera}}]{Wilhelmsen2015}%
  \BibitemOpen
  \bibfield  {author} {\bibinfo {author} {\bibfnamefont {{\O}.}~\bibnamefont
  {Wilhelmsen}}, \bibinfo {author} {\bibfnamefont {D.}~\bibnamefont {Bedeaux}},
  \ and\ \bibinfo {author} {\bibfnamefont {D.}~\bibnamefont {Reguera}},\ }\href
  {\doibase 10.1063/1.4919689} {\bibfield  {journal} {\bibinfo  {journal} {J.
  Chem. Phys.}\ }\textbf {\bibinfo {volume} {142}},\ \bibinfo {pages} {171103}
  (\bibinfo {year} {2015})}\BibitemShut {NoStop}%
\bibitem [{\citenamefont {Marin}\ \emph {et~al.}(2019)\citenamefont {Marin},
  \citenamefont {Alesker}, \citenamefont {Guttman}, \citenamefont {Gershinsky},
  \citenamefont {Edri}, \citenamefont {Shpaisman}, \citenamefont {Guerra},
  \citenamefont {Zitoun}, \citenamefont {Deutsch},\ and\ \citenamefont
  {Sloutskin}}]{Marin2019}%
  \BibitemOpen
  \bibfield  {author} {\bibinfo {author} {\bibfnamefont {O.}~\bibnamefont
  {Marin}}, \bibinfo {author} {\bibfnamefont {M.}~\bibnamefont {Alesker}},
  \bibinfo {author} {\bibfnamefont {S.}~\bibnamefont {Guttman}}, \bibinfo
  {author} {\bibfnamefont {G.}~\bibnamefont {Gershinsky}}, \bibinfo {author}
  {\bibfnamefont {E.}~\bibnamefont {Edri}}, \bibinfo {author} {\bibfnamefont
  {H.}~\bibnamefont {Shpaisman}}, \bibinfo {author} {\bibfnamefont {R.~E.}\
  \bibnamefont {Guerra}}, \bibinfo {author} {\bibfnamefont {D.}~\bibnamefont
  {Zitoun}}, \bibinfo {author} {\bibfnamefont {M.}~\bibnamefont {Deutsch}}, \
  and\ \bibinfo {author} {\bibfnamefont {E.}~\bibnamefont {Sloutskin}},\ }\href
  {\doibase 10.1016/j.jcis.2018.11.111} {\bibfield  {journal} {\bibinfo
  {journal} {J. Colloid. Interface Sci.}\ }\textbf {\bibinfo {volume} {538}},\
  \bibinfo {pages} {541} (\bibinfo {year} {2019})}\BibitemShut {NoStop}%
\bibitem [{\citenamefont {Bausch}\ \emph {et~al.}(2003)\citenamefont {Bausch},
  \citenamefont {Bowick}, \citenamefont {Cacciuto}, \citenamefont {Dinsmore},
  \citenamefont {Hsu}, \citenamefont {Nelson}, \citenamefont {Nikolaides},
  \citenamefont {Travesset},\ and\ \citenamefont {Weitz}}]{Bausch2003}%
  \BibitemOpen
  \bibfield  {author} {\bibinfo {author} {\bibfnamefont {A.~R.}\ \bibnamefont
  {Bausch}}, \bibinfo {author} {\bibfnamefont {M.~J.}\ \bibnamefont {Bowick}},
  \bibinfo {author} {\bibfnamefont {A.}~\bibnamefont {Cacciuto}}, \bibinfo
  {author} {\bibfnamefont {A.~D.}\ \bibnamefont {Dinsmore}}, \bibinfo {author}
  {\bibfnamefont {M.~F.}\ \bibnamefont {Hsu}}, \bibinfo {author} {\bibfnamefont
  {D.~R.}\ \bibnamefont {Nelson}}, \bibinfo {author} {\bibfnamefont {M.~G.}\
  \bibnamefont {Nikolaides}}, \bibinfo {author} {\bibfnamefont
  {A.}~\bibnamefont {Travesset}}, \ and\ \bibinfo {author} {\bibfnamefont
  {D.~A.}\ \bibnamefont {Weitz}},\ }\href {\doibase 10.1126/science.1081160}
  {\bibfield  {journal} {\bibinfo  {journal} {Science}\ }\textbf {\bibinfo
  {volume} {299}},\ \bibinfo {pages} {1716} (\bibinfo {year}
  {2003})}\BibitemShut {NoStop}%
\bibitem [{\citenamefont {Garc\'{\i}a-Aguilar}\ \emph
  {et~al.}(2020)\citenamefont {Garc\'{\i}a-Aguilar}, \citenamefont {Fonda},\
  and\ \citenamefont {Giomi}}]{GarciaAguilar2020}%
  \BibitemOpen
  \bibfield  {author} {\bibinfo {author} {\bibfnamefont {I.}~\bibnamefont
  {Garc\'{\i}a-Aguilar}}, \bibinfo {author} {\bibfnamefont {P.}~\bibnamefont
  {Fonda}}, \ and\ \bibinfo {author} {\bibfnamefont {L.}~\bibnamefont
  {Giomi}},\ }\href {\doibase 10.1103/PhysRevE.101.063005} {\bibfield
  {journal} {\bibinfo  {journal} {Phys. Rev. E}\ }\textbf {\bibinfo {volume}
  {101}},\ \bibinfo {pages} {063005} (\bibinfo {year} {2020})}\BibitemShut
  {NoStop}%
\bibitem [{\citenamefont {Sullivan}(2008)}]{Sullivan2008}%
  \BibitemOpen
  \bibfield  {author} {\bibinfo {author} {\bibfnamefont {J.~M.}\ \bibnamefont
  {Sullivan}},\ }\enquote {\bibinfo {title} {Curvatures of smooth and discrete
  surfaces},}\ in\ \href {\doibase 10.1007/978-3-7643-8621-4_9} {\emph
  {\bibinfo {booktitle} {Discrete Differential Geometry}}},\ \bibinfo {editor}
  {edited by\ \bibinfo {editor} {\bibfnamefont {A.~I.}\ \bibnamefont
  {Bobenko}}, \bibinfo {editor} {\bibfnamefont {J.~M.}\ \bibnamefont
  {Sullivan}}, \bibinfo {editor} {\bibfnamefont {P.}~\bibnamefont {Schröder}},
  \ and\ \bibinfo {editor} {\bibfnamefont {G.~M.}\ \bibnamefont {Ziegler}}}\
  (\bibinfo  {publisher} {Birkhäuser Basel},\ \bibinfo {address} {Basel},\
  \bibinfo {year} {2008})\ p.\ \bibinfo {pages} {175–188}\BibitemShut
  {NoStop}%
\bibitem [{\citenamefont {Brakke}(1992)}]{SurfaceEvolver}%
  \BibitemOpen
  \bibfield  {author} {\bibinfo {author} {\bibfnamefont {K.~A.}\ \bibnamefont
  {Brakke}},\ }\href {\doibase 10.1080/10586458.1992.10504253} {\bibfield
  {journal} {\bibinfo  {journal} {Experimental Mathematics}\ }\textbf {\bibinfo
  {volume} {1}},\ \bibinfo {pages} {141–165} (\bibinfo {year}
  {1992})}\BibitemShut {NoStop}%
\bibitem [{\citenamefont {Ashcroft}\ and\ \citenamefont
  {Mermin}(1976)}]{AshcroftMermim}%
  \BibitemOpen
  \bibfield  {author} {\bibinfo {author} {\bibfnamefont {N.}~\bibnamefont
  {Ashcroft}}\ and\ \bibinfo {author} {\bibfnamefont {N.}~\bibnamefont
  {Mermin}},\ }\href@noop {} {\emph {\bibinfo {title} {Solid State Physics}}}\
  (\bibinfo  {publisher} {Holt, Rine Hart and Winston},\ \bibinfo {year}
  {1976})\BibitemShut {NoStop}%
\bibitem [{\citenamefont {Yefet}\ \emph {et~al.}(2014)\citenamefont {Yefet},
  \citenamefont {Sloutskin}, \citenamefont {Tamam}, \citenamefont {Sapir},
  \citenamefont {Deutsch},\ and\ \citenamefont {Ocko}}]{Yefet2014-2}%
  \BibitemOpen
  \bibfield  {author} {\bibinfo {author} {\bibfnamefont {S.}~\bibnamefont
  {Yefet}}, \bibinfo {author} {\bibfnamefont {E.}~\bibnamefont {Sloutskin}},
  \bibinfo {author} {\bibfnamefont {L.}~\bibnamefont {Tamam}}, \bibinfo
  {author} {\bibfnamefont {Z.}~\bibnamefont {Sapir}}, \bibinfo {author}
  {\bibfnamefont {M.}~\bibnamefont {Deutsch}}, \ and\ \bibinfo {author}
  {\bibfnamefont {B.~M.}\ \bibnamefont {Ocko}},\ }\href {\doibase
  10.1021/la501589t} {\bibfield  {journal} {\bibinfo  {journal} {Langmuir}\
  }\textbf {\bibinfo {volume} {30}},\ \bibinfo {pages} {8010} (\bibinfo {year}
  {2014})}\BibitemShut {NoStop}%
\bibitem [{\citenamefont {Barker}(1963)}]{Barker1963}%
  \BibitemOpen
  \bibfield  {author} {\bibinfo {author} {\bibfnamefont {R.~E.}\ \bibnamefont
  {Barker}},\ }\href {\doibase 10.1063/1.1729049} {\bibfield  {journal}
  {\bibinfo  {journal} {J. Appl. Phys.}\ }\textbf {\bibinfo {volume} {34}},\
  \bibinfo {pages} {107} (\bibinfo {year} {1963})}\BibitemShut {NoStop}%
\end{thebibliography}%

\end{document}